\documentclass{PoS} 
\usepackage{graphicx}

\title{$B^0_s$ and $B^0$ Mixing in the Standard Model and Beyond:  A Progress Report} 

\ShortTitle{$B^0_s$ and $B^0$ Mixing in the Standard Model and Beyond:  A Progress Report} 

\author{%
\speaker{C.~Bouchard}\thanks{Supported in part by a Fermilab Fellowship in Theoretical Physics.}$^{\ \ a,c}$,
A.X.~El-Khadra$^{\ a}$,
E.D.~Freeland$^{\ a,b}$,
E.~G\'amiz$^{\ c}$, and
A.S.~Kronfeld$^{\ c}$
\\ \\
\llap{$^a$}Physics Department, University of Illinois, Urbana, IL  61801, USA \\
\llap{$^b$}Department of Physics, Washington University, St.~Louis, MO  63130, USA \\
\llap{$^c$}Theoretical Physics Department, Fermi National Accelerator Laboratory\thanks{Operated by 
Fermi Research Alliance, LLC, under Contract No.~DE-AC02-07CH11359
with the United States Department of Energy.},
~Batavia, IL  60510, USA}

\author{Fermilab Lattice and MILC Collaborations\\
        E-mail: \email{cbouchrd@illinois.edu}}

\abstract{We give a progress report on the calculation of $B$ meson mixing matrix elements, focusing on contributions that could arise beyond the Standard Model. The calculation uses asqtad (light quark) and Fermilab (heavy quark) valence actions and MILC ensembles with 2+1 flavors of asqtad sea quarks.  We report preliminary $B^0_s$ fit results, at a lattice spacing of 0.12 fm, for the SUSY basis of effective four-quark mixing operators and include an estimate for the final error budget.} 

\FullConference{The XXVIII International Symposium on Lattice Field Theory\\ 
June 14-19,2010\\ 
Villasimius, Sardinia Italy} 

\begin{document} 

\section{Introduction}

\subsection{Motivation}
The absence of tree-level flavor-changing neutral currents in the SM means its contributions to neutral meson mixing begin at the loop level.  The $B^0_s$ $\left(B^0\right)$ mixing amplitude is further suppressed by a combination of the GIM mechanism and the CKM matrix element $|V_{ts}|^2 \sim 2\times10^{-3}$ $\left(|V_{td}|^2 \sim 6\times10^{-5}\right)$, opening the door for observable new physics effects~\cite{Buras:2009}.

A number of recent analyses have revealed hints of new physics in $B^0_s$ and $B^0$ mixing.  A $(2-3)\sigma$ tension in the Unitarity Triangle can be explained by new physics in $\Delta M_s / \Delta M_d$~\cite{Lunghi:Laiho:Lenz}.  The UTfit collaboration performed a simultaneous analysis of multiple experimental results, finding $\sim 3 \sigma$ evidence of new physics in $B^0_s$ mixing~\cite{Bona:2009}.  $D\emptyset$'s recent report of an anomalous like-sign dimuon charge asymmetry in semileptonic b-hadron decay deviates from the SM prediction by $\sim 3 \sigma$, providing additional evidence of new physics in $B^0_s$ mixing~\cite{Abazov:2010,Lenz:2007}.

Disagreement between SM mixing predictions and experiment motivates increased precision in the calculation of SM hadronic mixing matrix elements.  These calculations have been performed on the lattice, with $2+1$ dynamical sea quarks, to a $(3-4)\%$ precision~\cite{Gamiz:Witzel,Evans:2009}, but further refinements are needed to sort out tension with experiment. To permit mixing predictions from new physics models, a knowledge of the possible BSM hadronic mixing matrix elements is required.  Pioneering work~\cite{Gimenez:2000,Becirevic:2001}, utilizing the quenched approximation and static limit of HQET, resulted in a quoted $\sim 10\%$ precision.  We aim to improve upon this.

\subsection{The Role of Lattice QCD}
Whether considering SM or beyond, the disparate scales of hadronization, $\mathcal{O}(500\ \textrm{MeV})$, and the underlying flavor-changing physics\footnote{In the SM this is the electroweak scale, $\mathcal{O}(100\ \textrm{GeV})$, and in new physics models is typically larger.} results in a factorization of the physics at the two scales.  For example, the SM expression for the oscillation frequency of the $B^0_q \leftrightarrow \overline{B^0_q}$ transition~\cite{Buras:1990},
\begin{equation}
\left(\Delta M_q \right)_{SM} = \left( \frac{G_F^2 M_W^2 S_0}{4 \pi^2 M_{B_q}} \right)\eta_B(\mu)\  \left| V_{tb} V_{tq}^* \right|^2\ \langle B_q^0|O(\mu)|\overline{B_q^0}\rangle,
\end{equation}
separates the low energy physics of hadronization, characterized by the mixing hadronic matrix element $\langle B_q^0|O|\overline{B_q^0}\rangle$, from the details of the flavor-changing interactions of the SM.  The SM plays a role in $\langle B_q^0|O|\overline{B_q^0}\rangle$ by restricting the structure of the effective four-quark operator $O$.
For a generic underlying theory we can write
\begin{equation}
\Delta M_q = \sum_i\ c_i(\mu)\ \langle B_q^0|O_i(\mu)|\overline{B_q^0}\rangle,
\end{equation}
where the short distance Wilson coefficients, $c_i$, contain the details of the underlying theory and are generally perturbatively calculable.  The $O_i$ are all effective, four-quark, $\Delta B =2$ ($B$ is the bottom quark flavor quantum number) operators allowed by the theory.

Lattice QCD permits calculation of the non-perturbative, purely hadronic quantities $\langle B_q^0|O_i|\overline{B_q^0}\rangle$. 

\section{Calculation} 

\subsection{Generating Data}
\begin{figure}[t]
	\begin{center}
	\scalebox{0.75}{\includegraphics[angle=0,width=1\textwidth]{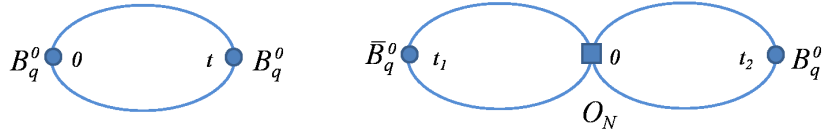}}
	\caption{(\textit{left}) Two- and (\textit{right}) three-point correlation functions in the meson rest frame.  The mixing operator is placed at the origin in the three-point correlation function.}
	\label{fig:correlators}
	\end{center}
\end{figure}

\begin{figure}[t]
\vspace{-0.35in}
\centering
{\scalebox{1.2}{\includegraphics[angle=0,width=0.41\textwidth]{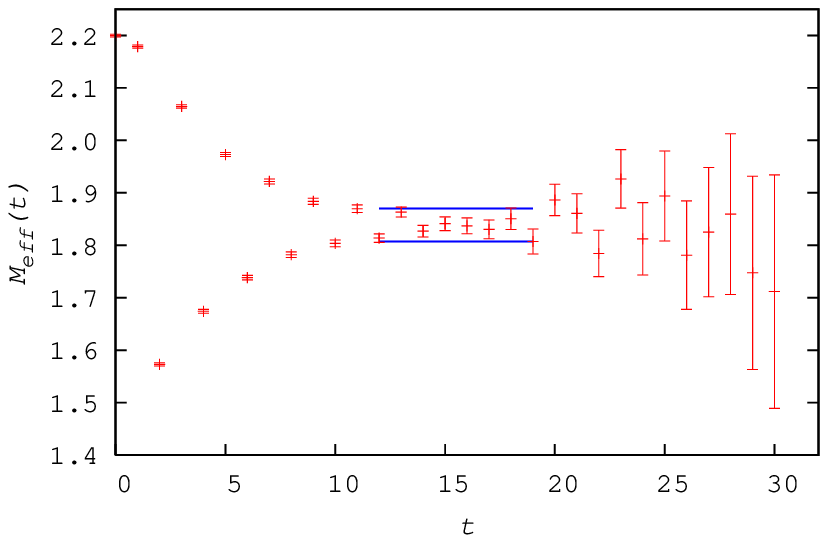}}}
{\scalebox{1.2}{\includegraphics[angle=0,width=0.41\textwidth]{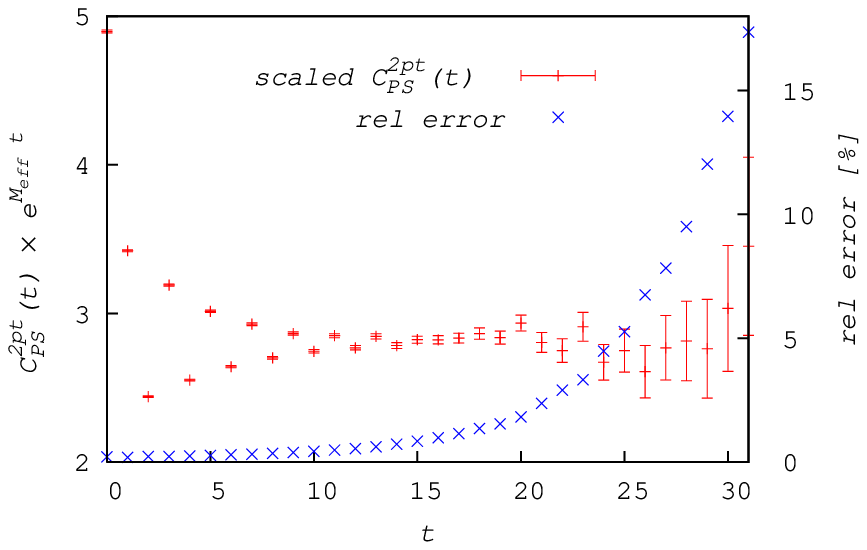}}}
\caption{\label{fig-2ptdata} (\textit{left}) $B_s$ effective mass plot and (\textit{right}) scaled two-point pseudo-scalar correlation function on the $a=0.12$ fm, $20\times 64^3$, $m_l/m_s=0.4$ ensemble.  Source and sink are 1S-smeared.}
\end{figure}

\begin{figure}[t]
\vspace{-0.2in}
\centering
{\scalebox{1.3}{\includegraphics[angle=0,width=0.41\textwidth]{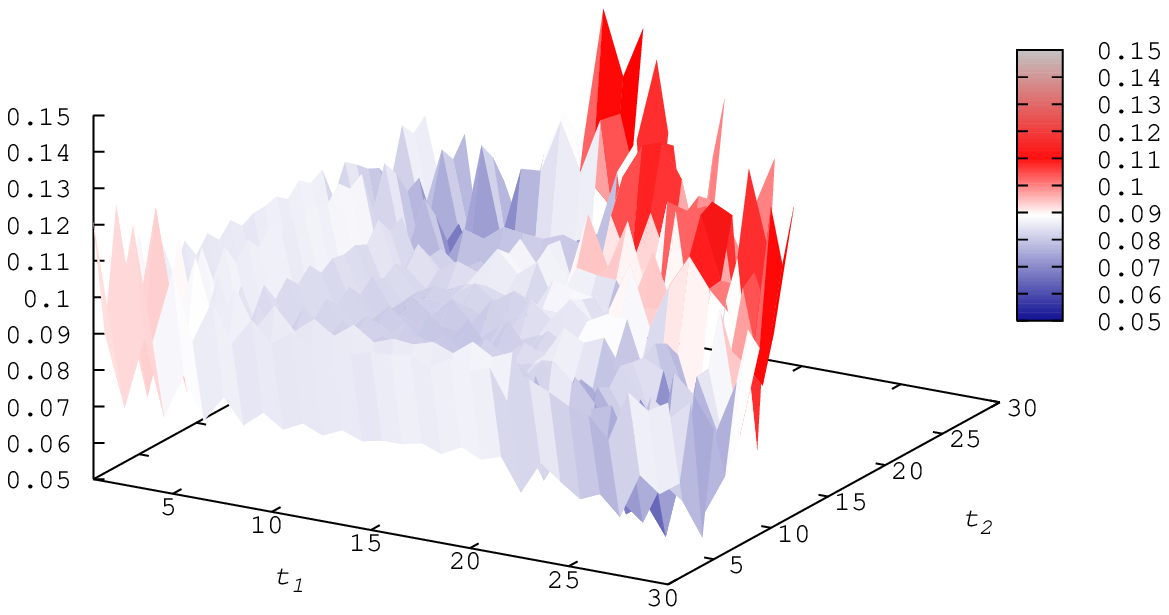}}}
\hspace{-0.5in}
{\scalebox{1.3}{\includegraphics[angle=0,width=0.41\textwidth]{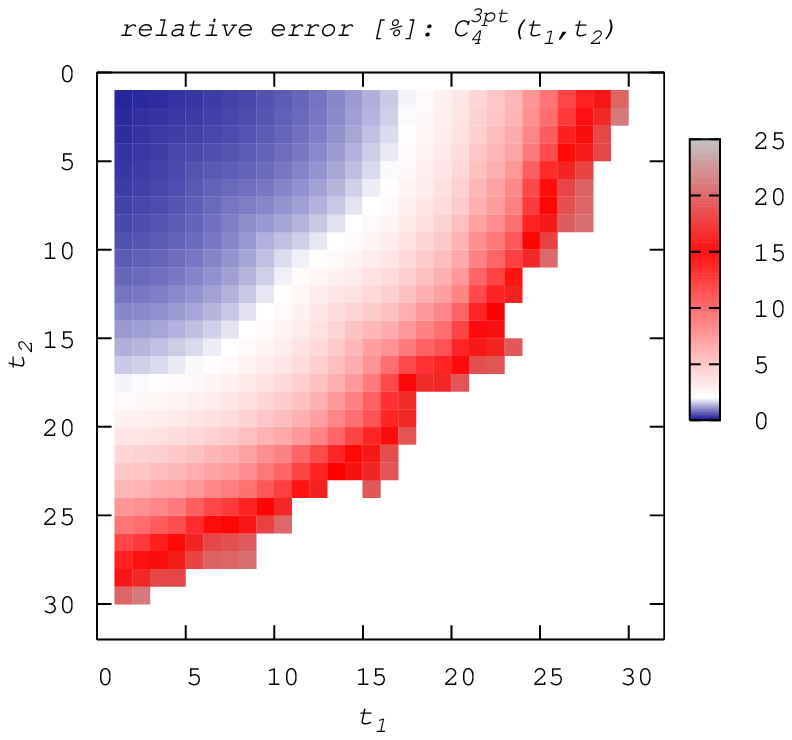}}}
\caption{\label{fig-3ptdata} (\textit{left}) The scaled $B_s$ three-point correlation function for mixing operator $O_4 = (\overline{b}^\alpha L\ q^\alpha) \ (\overline{b}^\beta R\ q^\beta)$ and (\textit{right}) its relative error on the $a=0.12$ fm, $20\times 64^3$, $m_l/m_s=0.4$ ensemble.  The heavy quark fields are 1S-smeared at the sinks.}
\end{figure}

The space of possible Lorentz-invariant, color singlet, four-quark mixing operators is spanned by a basis of mixing operators, referred to for historical reasons as the SUSY basis~\cite{Gabbiani:1996},
\begin{eqnarray}
\begin{array}{ll}
O_1 = ( \overline{b}^\alpha\gamma_{\mu} P_L\ q^\alpha) \ (\overline{b}^\beta\gamma_{\mu} P_L\ q^\beta),\ \ \ &\ \ \ O_2 = (\overline{b}^\alpha P_L\ q^\alpha) \ (\overline{b}^\beta P_L\ q^\beta), \\
O_3 = (\overline{b}^\alpha P_L\ q^\beta) \ (\overline{b}^\beta P_L\ q^\alpha),\ \ \ &\ \ \ O_4 = (\overline{b}^\alpha P_L\ q^\alpha) \ (\overline{b}^\beta P_R\ q^\beta),\ \textrm{and} \\
O_5 = (\overline{b}^\alpha P_L\ q^\beta) \ (\overline{b}^\beta P_R\ q^\alpha),\ \ \ &
\end{array}
\end{eqnarray}
listed here with greek color and suppressed spin indices.  Of 20 potential mixing operators, 12 can be eliminated by Fierz transformation and three by parity symmetry of QCD.  We're studying the possibility of using these extra operators to effectively increase statistics.  For each $O_N$ we write the mixing three-point correlation function as a time-ordered VEV of interpolating and mixing operators (a similar, if simpler, process is used to construct two-point correlation functions): 
\begin{eqnarray}
C^{3pt}_N(t_1,t_1) &=& \sum_{\vec{x_1},\vec{x_2}} \langle T\{ (\overline{q}\gamma_5 b)_{\vec{x}_2,t_2} (O_N)_{\vec{0},0} (\overline{q}\gamma_5 b)_{\vec{x}_1,t_1}\}\rangle, \\
C^{2pt}_{PS}(t) &=& \sum_{\vec{x}} \langle T\{ (\overline{q}\gamma_5 b)_{\vec{x},t} (q\gamma_5 \overline{b})_{\vec{0},0}\}\rangle.
\label{eqn-corr2}
\end{eqnarray}
Wick contraction yields products of quark propagators with time ordering ensured by imposing $t_2>0>t_1$ and $t>0$.  Heavy (light) quark propagators are obtained by inverting the Fermilab (asqtad) action on MILC gauge configurations with $2+1$ dynamical asqtad sea quarks~\cite{Bazavov:2010}.  A summary of gauge field configurations used (or planned for use) is given in Table~\ref{tab-ens}. 
\renewcommand*\arraystretch{1.2}
\begin{table}[b]
\begin{center}
\begin{tabular}{llllll}   \hline\hline                                       
$\approx a$ (fm) \hspace{0.1in}&\hspace{0.1in} $L^3 \times T$  \hspace{0.1in}&\hspace{0.1in} $\beta$ \hspace{0.1in}&\hspace{0.1in} $m_l/m_s$    \hspace{0.1in}&\hspace{0.1in} $m_{\pi}L$ \hspace{0.1in}&\hspace{0.1in} $N_c\times N_t$ \\  \hline
0.12             \hspace{0.1in}&\hspace{0.1in} $24^3\times$64  \hspace{0.1in}&\hspace{0.1in} 6.760   \hspace{0.1in}&\hspace{0.1in} 0.1          \hspace{0.1in}&\hspace{0.1in} 3.84 \hspace{0.1in}&\hspace{0.1in} 2099 $\times$ 4 \vspace{-0.15cm} \\ 
0.12             \hspace{0.1in}&\hspace{0.1in} $20^3\times$64  \hspace{0.1in}&\hspace{0.1in} 6.760   \hspace{0.1in}&\hspace{0.1in} 0.14         \hspace{0.1in}&\hspace{0.1in} 3.78 \hspace{0.1in}&\hspace{0.1in} 2110 $\times$ 4 \vspace{-0.15cm} \\ 
0.12             \hspace{0.1in}&\hspace{0.1in} $20^3\times$64  \hspace{0.1in}&\hspace{0.1in} 6.760   \hspace{0.1in}&\hspace{0.1in} 0.2          \hspace{0.1in}&\hspace{0.1in} 4.48 \hspace{0.1in}&\hspace{0.1in} 2259 $\times$ 4 \vspace{-0.15cm} \\ 
0.12             \hspace{0.1in}&\hspace{0.1in} $20^3\times$64  \hspace{0.1in}&\hspace{0.1in} 6.790   \hspace{0.1in}&\hspace{0.1in} 0.4          \hspace{0.1in}&\hspace{0.1in} 6.22 \hspace{0.1in}&\hspace{0.1in} 2052 $\times$ 4 \vspace{-0.15cm} \\
 & & &   \vspace{-0.2in} \\
0.09             \hspace{0.1in}&\hspace{0.1in} $40^3\times$96  \hspace{0.1in}&\hspace{0.1in} 7.080   \hspace{0.1in}&\hspace{0.1in} 0.1          \hspace{0.1in}&\hspace{0.1in} 4.21 \hspace{0.1in}&\hspace{0.1in} 1015 $\times$ 4 \vspace{-0.15cm}       \\ 
0.09             \hspace{0.1in}&\hspace{0.1in} $32^3\times$96  \hspace{0.1in}&\hspace{0.1in} 7.085   \hspace{0.1in}&\hspace{0.1in} 0.15         \hspace{0.1in}&\hspace{0.1in} 4.11 \hspace{0.1in}&\hspace{0.1in} 984 $\times$ 4 \vspace{-0.15cm}       \\ 
0.09             \hspace{0.1in}&\hspace{0.1in} $28^3\times$96  \hspace{0.1in}&\hspace{0.1in} 7.090   \hspace{0.1in}&\hspace{0.1in} 0.2          \hspace{0.1in}&\hspace{0.1in} 4.14 \hspace{0.1in}&\hspace{0.1in} 1931 $\times$ 4 \vspace{-0.15cm}       \\ 
0.09             \hspace{0.1in}&\hspace{0.1in} $28^3\times$96  \hspace{0.1in}&\hspace{0.1in} 7.110   \hspace{0.1in}&\hspace{0.1in} 0.4          \hspace{0.1in}&\hspace{0.1in} 5.78 \hspace{0.1in}&\hspace{0.1in} 1996 $\times$ 4 \vspace{-0.15cm}       \\ 
 & & &   \vspace{-0.2in} \\
0.06             \hspace{0.1in}&\hspace{0.1in} $64^3\times$144  \hspace{0.1in}&\hspace{0.1in} 7.460  \hspace{0.1in}&\hspace{0.1in} 0.1          \hspace{0.1in}&\hspace{0.1in} 4.27 \hspace{0.1in}&\hspace{0.1in} 826 $\times$ $N_t$ \vspace{-0.15cm}        \\ 
0.06             \hspace{0.1in}&\hspace{0.1in} $56^3\times$144  \hspace{0.1in}&\hspace{0.1in} 7.465  \hspace{0.1in}&\hspace{0.1in} 0.14         \hspace{0.1in}&\hspace{0.1in} 4.39 \hspace{0.1in}&\hspace{0.1in} 800 $\times$ $N_t$ \vspace{-0.15cm}        \\ 
0.06             \hspace{0.1in}&\hspace{0.1in} $48^3\times$144  \hspace{0.1in}&\hspace{0.1in} 7.470  \hspace{0.1in}&\hspace{0.1in} 0.2          \hspace{0.1in}&\hspace{0.1in} 4.49 \hspace{0.1in}&\hspace{0.1in} 668 $\times$ $N_t$ \vspace{-0.15cm}        \\ 
0.06             \hspace{0.1in}&\hspace{0.1in} $48^3\times$144  \hspace{0.1in}&\hspace{0.1in} 7.480  \hspace{0.1in}&\hspace{0.1in} 0.4          \hspace{0.1in}&\hspace{0.1in} 6.33 \hspace{0.1in}&\hspace{0.1in} 668 $\times$ $N_t$ \vspace{-0.15cm}        \\
 & & &   \vspace{-0.2in} \\
0.045           \hspace{0.1in}&\hspace{0.1in}  $64^3\times$192  \hspace{0.1in}&\hspace{0.1in} 7.810  \hspace{0.1in}&\hspace{0.1in} 0.2          \hspace{0.1in}&\hspace{0.1in} 4.56 \hspace{0.1in}&\hspace{0.1in} 860 $\times$ $N_t$  \\  \hline\hline
\end{tabular}
\end{center}
\caption{MILC ensembles~\cite{Bazavov:2010} planned for use in this study.  $N_c\times N_t$ is the number of configurations and source times (not yet determined if unspecified).  This report includes results for the $a=0.12$ fm ensembles.}
\label{tab-ens}
\end{table}
We work in the meson rest frame by Fourier transforming the correlation functions and setting $\vec{p}=0$, leaving correlation functions that depend only on time (depicted in Fig.~\ref{fig:correlators}).  
In the three-point correlation functions, the heavy quark fields in the mixing operator are improved to remove a discretization error~\cite{El-Khadra:1997} and 1S-smeared at the sinks to increase ground state overlap.  Heavy quark fields in the two-point correlation functions are 1S-smeared at the source and sink.  Figs.~\ref{fig-2ptdata} and~\ref{fig-3ptdata} show plots of two- and three-point correlation function data generated in this way.  All data and fit results shown are for the $B_s$ meson ($\kappa=0.0918$ and $ma=0.0349$).


\subsection{Fitting}

We build fit functions by decomposing the two- and three-point correlation functions of Eqs.~(2.2 and \ref{eqn-corr2}) in a basis of energy eigenstates, giving an infinite sum of exponentials
\begin{eqnarray}
C^{2pt}_{PS}(t) &=& \sum_{n=0}^{\infty} \frac{ |Z_n|^2}{2E_n}(-)^{n(t+1)}\ \left( e^{-E_n t} + e^{-E_n (T-t)} \right) \\
C^{3pt}_N(t_1,t_1) &=& \sum_{n,m=0}^{\infty} \frac{\langle n| O_N |m\rangle Z_n^{\dagger}Z_m}{4E_nE_m}(-)^{n(t_1+1)+m(t_2+1)}\ \left( e^{-E_n|t_1|} + e^{-E_n(T-|t_1|)} \right) \left(e^{-E_mt_2} + e^{-E_m(T-t_2)} \right) \nonumber
\label{eqn-corrsum}
\end{eqnarray}
where $Z_n\equiv\langle n|\overline{q}\gamma_5 b\rangle$.  Oscillating opposite parity states, a result of staggered light valence quarks~\cite{Wingate:2003}, and the effect of periodic boundary conditions are accounted for in Eq.~(2.4).  In practice, we limit the time range of data included in the fit $(t_{\mathrm{min}}\leq t \leq t_{\mathrm{max}})$ and truncate the sums $(\sum_{n=0}^{N^{2pt}-1}\textrm{ and } \sum_{n,m=0}^{N^{3pt}-1})$, where $N^{2,3pt}$ is the number of states used in the fit.

Fits using data at short times must account for increased excited state contributions by including an adequate number of states.  Despite added difficulty, the relatively clean signal in the data at short times may make it desirable to include them in the fit.  We accomplish this using a Bayesian fitting routine~\cite{Lepage:2001,Lepage:2008} and a systematic procedure to select $N$, $t_{\mathrm{min}}$ and $t_{\mathrm{max}}$.  We are able to achieve consistent and stable fits, with a suitable choice of time range, for $N^{2,3pt}=2,4$ and 6.
\begin{figure}[t]
\centering
{\scalebox{1.28}
{\includegraphics[angle=0,width=0.39\textwidth]{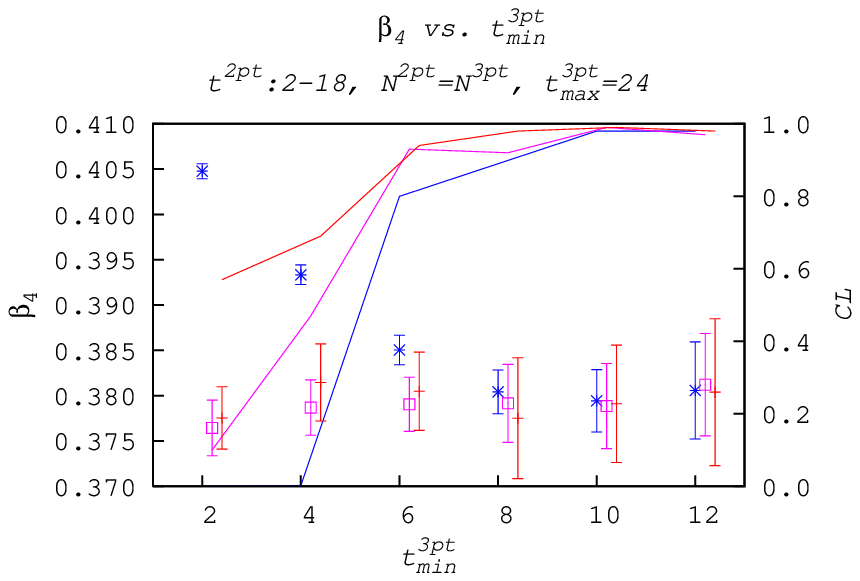}}}
\hspace{-0.09in}
{\scalebox{1.28}{\includegraphics[angle=0,width=0.39\textwidth]{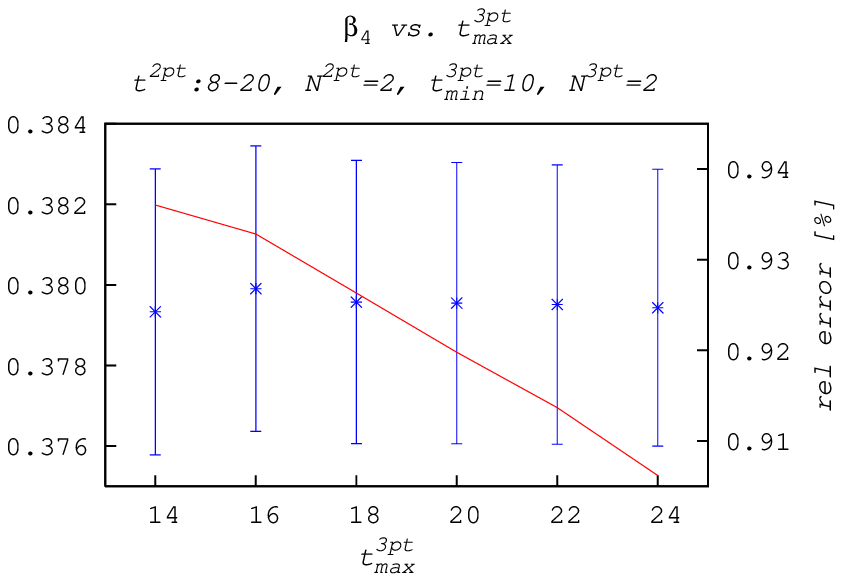}}}
\caption{\label{fig-3ptfits}Simultaneous fits of $C^{2pt}_{PS}$ and $C^{3pt}_4$ on the $a=0.12$ fm, $20\times 64^3$, $m_l/m_s=0.4$ ensemble for the $B_s$ mixing parameter $\beta_4$, defined in Sect. 3. (\textit{left}) Fits vs. $t_{\mathrm{min}}$ (shown for $t_{\mathrm{max}}=24$) reveal a common plateau for $N^{3pt} = N^{2pt} = 2$ (blue burst), 4 (pink square) and 6 (red dash) fits.  Solid lines indicate confidence levels for the fits.  (\textit{right}) We plot a representative fit from the plateau ($N^{3pt} = 2$ and $t_{\mathrm{min}}^{3pt}=10$) to demonstrate stability with respect to $t_{\mathrm{max}}^{3pt}$.  Fits are blue bursts with error bars and the solid red line is the relative error.}
\end{figure}

From scaled correlation functions we determine time ranges to study.  For $t_{\mathrm{min}}$ we generally consider from $t_{\mathrm{min}} = 2$ until excited state contributions have significantly decreased.  Though increasing $t_{\mathrm{max}}$ utilizes more data, it also introduces an increasing level of noise and can lead to an underdetermined covariance matrix.  For the two-point correlation function of Fig.~\ref{fig-2ptdata} we considered $t_{\mathrm{min}}^{2pt}=2,4,...,12$ and $t_{\mathrm{max}}^{2pt}=12,14,...,24$, for the three-point correlation function of Fig.~\ref{fig-3ptdata}, $t_{\mathrm{min}}^{3pt}=2,4,...,12$ and $t_{\mathrm{max}}^{3pt}=14,16,...,24$, and in each case $N^{2,3pt}=2,4,6$.

We fit for each combination of $N^{2pt}$, $t_{\mathrm{min}}^{2pt}$ and $t_{\mathrm{max}}^{2pt}$ and select a representative fit from the common plateau, ensuring stability with respect to our choice of $N^{2pt}$, $t_{\mathrm{min}}^{2pt}$ and $t_{\mathrm{max}}^{2pt}$.  Then, fixing the two-point fit parameters, we repeat the procedure for a simultaneous fit of the two- and three-point correlation functions.  Fig.~\ref{fig-3ptfits} demonstrates the stability of the simultaneous fits.
\subsection{Initial Results}
\label{sec-results}
Table~\ref{tab-results} lists preliminary fit results for the $B_s$ mixing parameters, $\beta_N$, defined by $\langle B_q^0|O_N|\overline{B_q^0}\rangle={\mathcal C}_N M_{B_q} \beta_N^2\ \ ({\mathcal C}_1=2/3, {\mathcal C}_2=-5/12, {\mathcal C}_3=1/12, {\mathcal C}_4=1/2,\ \mathrm{and}\ {\mathcal C}_5=1/6)$.
\renewcommand*\arraystretch{1.2}
\begin{table}[h!]
\begin{center}
\begin{tabular}{rllll}   \hline\hline
$m_l/m_s=$ \hspace{0.1in}&\hspace{0.1in} 0.4  \hspace{0.1in}&\hspace{0.1in} 0.2 \hspace{0.1in}&\hspace{0.1in} 0.14  \hspace{0.1in}&\hspace{0.1in} 0.1 \\  \hline
$\beta_1 \times (r_1/a)^{3/2}=$ \hspace{0.1in}&\hspace{0.1in} 1.217(14) \hspace{0.1in}&\hspace{0.1in} 1.196(11) \hspace{0.1in}&\hspace{0.1in} 1.160(14) \hspace{0.1in}&\hspace{0.1in} 1.161(14) \vspace{-0.0cm} \\ 
$\beta_2 \times (r_1/a)^{3/2}=$ \hspace{0.1in}&\hspace{0.1in} 1.509(15) \hspace{0.1in}&\hspace{0.1in} 1.446(14) \hspace{0.1in}&\hspace{0.1in} 1.425(16) \hspace{0.1in}&\hspace{0.1in} 1.448(14) \vspace{-0.0cm} \\ 
$\beta_3 \times (r_1/a)^{3/2}=$ \hspace{0.1in}&\hspace{0.1in} 1.490(22) \hspace{0.1in}&\hspace{0.1in} 1.409(25) \hspace{0.1in}&\hspace{0.1in} 1.318(34) \hspace{0.1in}&\hspace{0.1in} 1.446(25) \vspace{-0.0cm} \\ 
$\beta_4 \times (r_1/a)^{3/2}=$ \hspace{0.1in}&\hspace{0.1in} 1.785(14) \hspace{0.1in}&\hspace{0.1in} 1.731(12) \hspace{0.1in}&\hspace{0.1in} 1.689(14) \hspace{0.1in}&\hspace{0.1in} 1.699(13) \vspace{-0.0cm} \\
$\beta_5 \times (r_1/a)^{3/2}=$ \hspace{0.1in}&\hspace{0.1in} 2.313(26) \hspace{0.1in}&\hspace{0.1in} 2.255(14) \hspace{0.1in}&\hspace{0.1in} 2.200(21) \hspace{0.1in}&\hspace{0.1in} 2.223(13) \\ \hline\hline
\end{tabular}
\end{center}
\caption{Preliminary fit results for $\beta_N$ on the $a=0.12$ fm ensembles.  Errors are statistical from the fit.}
\label{tab-results}
\end{table}

\section{Outlook}

We are extending $B^0_s$ fits to other lattice spacings and a range of valence masses, to include $B^0_d$.  We will use these fits in an extrapolation to physical light sea quark mass, the continuum, and light valence quark mass (and an interpolation to physical strange quark mass).  The continuum \cite{Detmold:2007} and staggered lattice~\cite{Laiho:2007} chiral perturbation theory has been worked out.  One-loop perturbative renormalization for the SM mixing operators exists and the BSM operator renormalization is expected to be a simple extension of this work.

\renewcommand*\arraystretch{1.2}
\begin{table}[h!]
\begin{center}
\begin{tabular}{llll}
\hline\hline
Source of Error [\%]                   \hspace{0.1in}&\hspace{0.1in} $\beta_1$ (\textit{Lattice 2009}) \hspace{0.1in}&\hspace{0.1in} Expected \hspace{0.1in}&\hspace{0.1in} Reference \\
\hline
statistical                            \hspace{0.1in}&\hspace{0.1in} 2.7                         \hspace{0.1in}&\hspace{0.1in} 1.2 \hspace{0.1in}&\hspace{0.1in} current work \vspace{-0.1cm} \\
scale ($r_1$)                           \hspace{0.1in}&\hspace{0.1in} 3.0                        \hspace{0.1in}&\hspace{0.1in} 1.0 \hspace{0.1in}&\hspace{0.1in} \cite{Bazavov:2009} \vspace{-0.1cm} \\
sea \& valence quark masses            \hspace{0.1in}&\hspace{0.1in} 0.3                         \hspace{0.1in}&\hspace{0.1in} 0.3 \hspace{0.1in}&\hspace{0.1in} \vspace{-0.1cm} \\
b-quark hopping parameter              \hspace{0.1in}&\hspace{0.1in} $\leq$ 0.5                  \hspace{0.1in}&\hspace{0.1in} 0.1 \hspace{0.1in}&\hspace{0.1in} \cite{Bailey:2010} \vspace{-0.1cm} \\
$\chi$PT $+$ light quark discretization \hspace{0.1in}&\hspace{0.1in} 0.4                        \hspace{0.1in}&\hspace{0.1in} ${<0.4}$ \hspace{0.1in}&\hspace{0.1in} * \vspace{-0.1cm} \\ 
$g_{B^*B\pi}$                          \hspace{0.1in}&\hspace{0.1in} 0.3                         \hspace{0.1in}&\hspace{0.1in} ${<0.3}$ \hspace{0.1in}&\hspace{0.1in} * \vspace{-0.1cm} \\
heavy quark discretization             \hspace{0.1in}&\hspace{0.1in} 2                           \hspace{0.1in}&\hspace{0.1in} $\sim 1.2$ \hspace{0.1in}&\hspace{0.1in} \cite{Bailey:2010} \vspace{-0.1cm} \\
matching (1-loop perturbation theory)  \hspace{0.1in}&\hspace{0.1in} $\sim$4                     \hspace{0.1in}&\hspace{0.1in} $\sim 2.5$ \hspace{0.1in}&\hspace{0.1in} \cite{Gamiz:2010} \vspace{-0.1cm} \\ 
finite volume effects                  \hspace{0.1in}&\hspace{0.1in} $\leq 0.5$                  \hspace{0.1in}&\hspace{0.1in} $<0.1$ \hspace{0.1in}&\hspace{0.1in} \cite{Bailey:2010} \\ 
\hline
Total                                  \hspace{0.1in}&\hspace{0.1in} 6.1                         \hspace{0.1in}&\hspace{0.1in} $\sim 3.2$ \\
\hline\hline
\end{tabular}
\end{center}
\caption{We estimate an error budget, for $\beta_1$, by way of comparison with~\cite{Evans:2009}.  The sources of expected improvements are listed in the Reference column.  ${}^*$We anticipate improvement from finer lattice spacings.}
\label{tab-error}
\end{table}
Accurately accounting for errors in the calculation is as important as achieving precision results.  Our naive statistical errors represent a (40-50)\% reduction relative to~\cite{Evans:2009}.  We are generating more robust estimates via the bootstrap method.  Correlator data exist for $a=0.12$, $0.09$ fm and are being generated for $a=0.06$, $0.045$ fm.  Table~\ref{tab-error} quotes statistical and systematic errors for $\beta_1$ from~\cite{Evans:2009}, with expected improvements from the use of $a=0.06$ fm data.  Analysis of $a=0.045$ fm data will further improve the error budget.

Computations for this work were carried out in part on facilities of the USQCD Collaboration, which are funded by the Office of Science of the U.S. Department of Energy. This work was supported in part by the U.S. Department of Energy under Grants No. DE-FG02-91ER40677 (A.X.K., C.B., E.D.F.), No. DEFG02-91ER40628 (E.D.F.); the National Science Foundation under Grant No. PHY-0555235 (E.D.F.).


\providecommand{\href}[2]{#2}\begingroup

\end{document}